# Exploring Gender Disparities in Bumble's Match Recommendations

*Research-in-Progress Paper*

**Ritvik Kalra**
International Institute of Information Technology Hyderabad, India,
ritvik.kalra@research.iiit.ac.in

**Pratham Gupta**
International Institute of Information Technology Hyderabad, India,
gupta.pratham@research.iiit.ac.in

**Ben Varghese**
International Institute of Information Technology Hyderabad, India,
ben.varghese@students.iiit.ac.in

**Nimmi Rangaswamy**
International Institute of Information Technology Hyderabad, India,
nimmi.rangaswamy@iiit.ac.in

## Abstract

We study bias and discrimination in the context of Bumble, an online dating platform in India. Drawing on research in AI fairness and inclusion studies we analyze algorithmic bias and their propensity to reproduce bias. We conducted an experiment to identify and address the presence of bias in the matching algorithms Bumble pushes to its users in the form of profiles for potential dates in the real world. Dating apps like Bumble utilize algorithms that learn from user data to make recommendations. Even if the algorithm does not have intentions or consciousness, it is a system created and maintained by humans. We attribute moral agency of such systems to be compositely derived from algorithmic mediations, the design and utilization of these platforms. Developers, designers, and operators of dating platforms thus have a moral obligation to mitigate biases in the algorithms to create inclusive platforms that affirm diverse social identities.

## Introduction

Anti-discrimination represents a critical area of study for researchers, policymakers, and designers of technical systems. We study bias and discrimination in the context of Bumble, an online dating platform among users in India. Drawing on AI fairness and inclusion studies research, we analyze algorithmic bias and their propensity to reproduce bias [Selbst et al., 2019]. Our study conducts an experiment on the Bumble platform to identify and address the presence of discrimination in the matching algorithms the platform pushes to its users in the form of profiles for potential dates in the real world. There has been a transformation in India's romantic and social interaction landscape, propelled by technological advancements and a shift in cultural attitudes. Dating applications have become increasingly popular for these kinds of interactions. The rise in internet users in India is projected to reach 1.3 billion in 2023 [Statista, 2023]. Online dating applications typically involve creating a profile on a dating website or app and then searching for and communicating with potential partners. These services use various technologies, such as location-based services and algorithms, to match users with compatible partners. Bumble's recommendation algorithm understands the dating preferences of the user and recommends other users based on these preferences. User preferences are based on learning a user's "swipe" behavior on the profiles displayed to them and how other profiles swipe on the user profile. Although the exact details of the recommendation engines are not public, we believe that Bumble's algorithm would be using an "Elo Score" which Tinder uses for their recommendation engine [Carr, 2016]. This score does not objectify

'attractiveness' of a user but measures their desirability and is based not just metrics like 'How many people swiped on you' but on multiple factors that are not available to the public eye.

The performance of the recommendation engines [ like in many machine learning models] depends on the representativeness and quality of their training data. It is highly likely that these algorithms push mainstream content and push diverse and minority ideas to the sidelines. A possible reason for these kinds of actions of recommendation algorithms could be because of not enough representation in the data. A lack of comprehensive diversity, hence, skews disproportionately towards prevailing societal preferences [Davidson et al., 2019]. This can narrow down the diversity of the content being pushed and hence reinforce pre-existing cultural biases. This can narrow the content exposure range and reinforce pre-existing cultural biases. Our paper aims to critically evaluate the recommendation algorithm used by Bumble in its profile suggestions as they affect interpersonal interactions and self-presentation. We specifically examine potential discrepancies and biases in the recommendations made to individuals identifying with dominant genders, namely men and women, in contrast to those of non-binary genders. Through this exploration, we aim to shed light on any inherent algorithmic inclinations and discuss their implications in the broader context of digital social interactions and inclusivity. This study is research in progress leading us to explore deeper questions on the power and influence of digital technologies on social life.

## Literature Review

The history of dating and dating applications in India has evolved significantly over time. Historically, methods to find love have included newspaper advertisements in the 18th century and 'radio-love' declarations in the 1900s. The rise of the internet in the late 1950s presented a new avenue to meet singles, and now digital screens dominate the dating landscape. [Forbes, 2020] Introduction of dating applications in India has had a significant effect on how intimacy and relationships are portrayed in the country. The impact can be considered analogous to how liberalization of the Indian economy brought upon a paradigm shift in the intimate lives of Indians. Historically, the idealization of the Indian middle-class wife viewed as virtuous champions of Indian morality, played a significant role in the construction of the Nationalist Identity of India. But liberalization brought upon exposure of the foreign media to the Indian masses, reflecting a modern form of the historical struggle, as Indian individuals and families navigate the challenges of integrating aspects of global culture while maintaining their distinct national and cultural identity. [Dell, 2005]

Dating applications are in a similar fashion a western influence that provide an alternative to the traditional route of arranged marriage, allowing people romantic relationships without the explicit end goal of marriage and individuals to explore their sexuality and desires not socially permissible until then. Introduction of platforms like Tinder, OkCupid and Bumble have basically shaped the online dating scene of India, while adapting and accommodating to the 'less casual approach' of Indians towards romantic relationships [Das, 2019]. The CMO of OkCupid once said "What makes India different is that we have a generation of young women who are changing things by saying, 'I want to decide who I'm with'" [Forbes, 2020] Homegrown Indian applications like TrulyMadly and Aisle are much more aligned with the Indian traditions, understanding the social norms and offer somewhat of a middle ground between casual dating and marriage [Forbes, 2020]. Bumble also had emerged as a champion of women empowerment, by making the women make the first move [Bumble].

Women in India have particularly embraced Bumble's features, making the first move over 15 million times and sending twice the number of messages compared to global averages [Hindustan Times, 2020]. Additionally, 32% of women users in India engage with more than one mode on the app, suggesting usage more than just dating [Hindustan Times, 2020]. A recent Bumble report suggests social media, music, food, literature, and content consumption are shaping GenZ's dating ideals in India. [Chronicle, 2023] The application has also predicted shifts in dating preferences, suggesting that 40% of single Indians will opt for virtual dating while placing equal importance on education, career choices, and compatibility. [MediaInfoline, 2023]

Introduction of Bumble has played an important part in showcasing LGBTQIA+ identities in India. Despite the government of India decriminalizing gay sex [Johar], social acceptance for the community is still inconsistent. Bumble has come up with guides for queer dating [Bumble] and along with its features like incognito mode, private detector, photo verification and pronoun selection have been instrumental steps in creation of Bumble as a safe space for LGBTQIA+ individuals [Bumble].

However, as recommendation algorithms become ubiquitous across various platforms, including shopping and social media, concerns about their potential for perpetuating social injustices are growing. Notable is the COMPAS algorithm, which had higher rates of misclassifying black defendants for violent recidivism risk than white defendants, with white recidivists often underestimated and black recidivists more likely to re-offend [Mattu, 2016]. In the domain of career recommendations [Lambrecht, 2019] ads for STEM Careers are shown more to men than women because showing to women was deemed more expensive by a cost optimization algorithm in ad delivery. XING, a job platform like LinkedIn, recommends more underqualified men to jobs than qualified women [Lahoti, 2019]. Attention has been paid to the complex interactions between technology and sociocultural dynamics in human-computer interaction (HCI) and social computing, particularly regarding power disparities and social justice concerns. Researchers have worked for positive community building by incorporating various studies from behavioral economics, computer science, and sociology towards commitment mechanisms, efficient participation control, and incentive structures in creating effective online communities [Hutson, 2018]. Their interdisciplinary approach highlights the relevance of basing online community design on well-established social psychology concepts, providing both practitioners and academics with priceless insights [Kraut, 2012]. In the domain of Social Justice, work has been done to suggest design practice that requires the commitments which are deemed necessary to combat social injustice, which include a commitment to reflexivity and personal ethics [Dombrowski, 2016].

Forms of interactions of user discovery pose a significant risk of sexual violence and other forms of harm for the LGBTQIA+ community on dating platforms. Design methods have suggested making dating applications more inclusive, safe, and comfortable for the people of the LGBTQIA+ community. Participatory design research suggests the inclusion of reduced visibility of profiles in newer geographical areas for women and other minorities, thereby envisioning dating apps to learn locations which are deemed safe. [Zytko, 2022].

Despite the growing popularity of Bumble, there is limited attention from the research community in India. A material-semiotic analysis on Bumble [Bivens, 2018] suggests the platform's infrastructure, gender-based safety features, and mechanisms are predominantly developed with heterosexuality dynamics in mind. Another interdisciplinary-triangulated-qualitative method to analyze the algorithms in dating applications concludes algorithms affect virtually every aspect of society and individual lives like an individual's mental health by creating feelings of self-doubt, the culture as the algorithm is shaping and getting shaped by the social norms and affects the social strata by reinforcing or mitigating social discrimination and gender inequality [Yang, 2021]. An interface analysis of Bumble argues its UI more aligned to the functional needs of matching profiles, rather than representing their identities imposing limitations on the user to self-represent on the platform leading to misunderstandings or misinterpretations [MacLeod, 2019]. On Tinder, beyond the above-mentioned studies [MacLeod 2019, Yang 2021], an analysis of profile data, to explore the behavior of men and women [Tyson, 2016] was conducted and found out that men are more non-selective about their matches in comparison to women, and women try to keep a decent interval between consecutive messages of their own.

We believe our study in reporting experiences of non-binary individuals using dating platforms contributes to a largely unexplored research field. The methodology adopted to understand these experiences is unlike previous research focusing primarily on the binary gender perspectives. As far as our knowledge permits us, we believe there are no studies investigating matching algorithms along the gender spectrum and in the context of Indian youth.

## Methodology

For the study, we recruited participants of different genders: two self-identified men, two self-identified women and one self-identified non-binary person. The participants were all active or past users of Bumble, which was important for our study to understand user experience on dating applications. All users belonged to the age group of 18 and 24. In India, coming out as a person of the LGBTQIA+ community is still a difficult choice, as social acceptance is deeply problematic. For the non-binary counterparts, Bumble asks you to fill in which gender category you want your profile to show as, i.e., either Man or Woman. We define these profiles as 'Non-Binary Men' and 'Non-Binary Women' respectively. The participants were asked to set up two Bumble profiles of themselves. Both profiles were precisely the same, with the only change being that one profile will show a man/woman while the other will be non-binary. For example, if a participant is a cisgender male, their other profile will show them as a non-binary male.

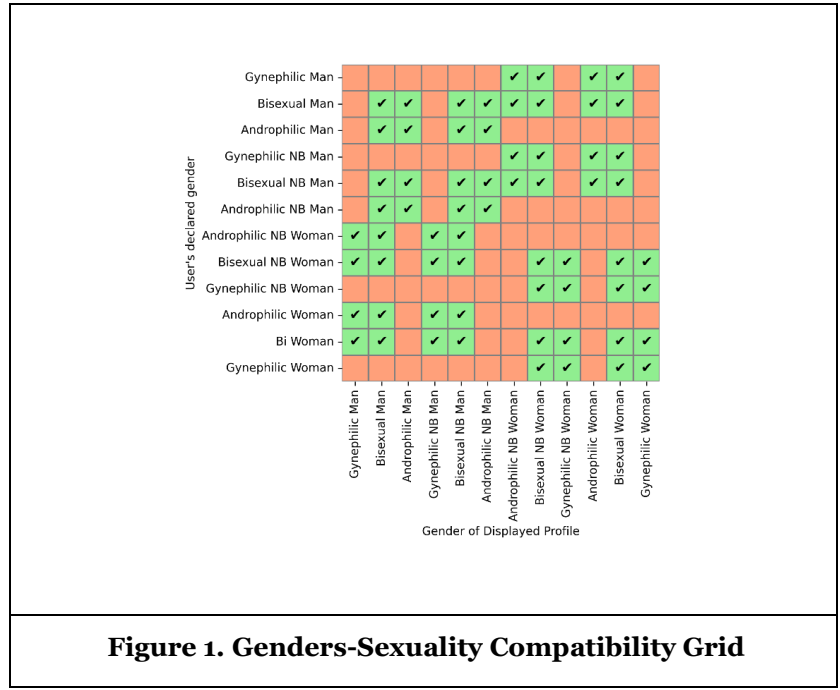

**Figure 1. Genders-Sexuality Compatibility Grid**

We have made a conscious choice of non-binary to be less specific about gender preference and also to adhere to the conventions of Bumble, which showcases the major genders as man, woman and non-Binary, though Bumble allows an individual to be more specific about their gender to represent themselves by giving options like Trans woman or man, Intersex man or woman, Bigender, Genderfluid and many more, with a total of 31 such genders [Bumble]. The exponential nature of studying the interaction between each of these genders prompted us to design our experiment with the three Bumble-specified major genders.

The participants were then asked to swipe at profiles Bumble showed them and capture profile details that were public. The details captured included Age, Gender, Work, Education, Location, Hometown, Height, Activity Level, Star Sign, Drinking, Smoking, Kids and Relationship preferences. All this data is chosen by the user who has created the profile to display on their own accord. They don't necessarily have to fill in any of these details and can choose to omit them, but the motivating factor provided by Bumble is to allow them to provide better potential suitors to the user. There are many confounding variables that come into play conducting a study like ours. The age of users, their technological literacy, how they approach dating applications can all affect the outcomes of this study. Besides, the location we choose to draw our dating profiles from can influence the algorithm- in our study we chose Mumbai as the location to set sample profiles. We chose Mumbai, because of the city's diverse user base who have varied dating preferences, high urban population and the tech-savvy population allowing for a more comprehensive study. Cultural and demographic factors could influence attitudes towards dating and the LGBTQIA+ community, thus affecting user experiences. Additionally, varying levels of social acceptance and stigma towards different gender identities in India could significantly affect user interactions. Being research-in-progress, we are hoping to deepen our understanding of confounding variables in the study.

The use of fictitious profiles for data collection on websites like Bumble raises several ethical issues, including informed consent, deception, and the risk of harm. Our methodology for collecting data is motivated by other peer-reviewed studies using fictitious Facebook profiles to gather data about university students [Mislove et al., 2010] and the use of fictitious profiles to understand the activity of users on Tinder [Tyson, 2016], and has been meticulously designed to adhere to the ethical standards. We only collect data that is regarded as public information [Elovici, 2014] from anyone on the platform. We refrained from instigating any user interaction between the participant and the profile displayed. If a match occurs, the user will be informed of the purposes of our research and consent will be actively sought before the study uses their data. Additionally, this research project has been thoroughly reviewed and granted clearance by the ethics board of our institute, ensuring compliance with both ethical and institutional guidelines.

We conducted interviews to get a deeper understanding of Bumble's recommendation algorithm and its outcomes for users. For this purpose, we decided to recruit people belonging to the same age group as the people that were recruited for the Bumble profiles, which are in the range of 19-24. This far we recruited two bisexual women, one bisexual man, one bisexual non-binary. It was a conscious decision to interview bisexual individuals, because of the high likelihood of them using the "open for everyone" filter in Bumble.

One of the first attributes that we wanted to analyze while trying to find a bias in the recommendations was that of gender, i.e., the ratios of the genders being shown to each of our profiles. We first counted the number of profiles of each gender that our users had observed and then normalized the count on the total number of profiles. We took these normalized counts of all genders shown to the participants and found the mean across all the participants with the same declared gender. This allowed us to get the distribution of profiles across genders for a generic user with a particular declared gender. The distribution of genders for a given participant's gender and their non-binary counterpart should have parity, considering that Bumble asks its non-binary users to put themselves in searches as Man or Woman. This is because they are seeking the same group of people, as seen in Figure 1. We observe skews in the gender distribution (Figure 2) between the dominant genders and their non-binary counterparts. Of the 2589 profiles swiped across all our subjects, the number of women that were displayed for our male profiles was 9.1% higher compared to our non-binary male profiles. On the other hand, our female profiles were shown 6.8% more men compared to our non-binary female profiles. We raise the question: If Bumble is not following the paradigm described in Fig 1 to showcase the profiles, then why is this? Another question that can be raised from this is, even if they are doing so, then why is this disparity among dominant genders and their counterparts?

In the current Bumble design, a nonbinary person is offered the unavoidable choice to show oneself in either men's or women's searches. The subsequent erasure of nonbinary identities that do not align with either man or woman exclusively at all (e.g., agender, bigender people) is another pitfall the design falls prey to.

Another key observation from this distribution is that of the skew in the number of men and women that are being displayed to the users. The ratio of men being shown to women is significantly high in general. This can be attributed to the gender ratio of users on Bumble, which corresponds to the mean percentage calculated for all profiles swiped through across all subjects (Women: 9.92%, Men: 88.12%, Non-binary: 1.96%). This skew becomes worse if we compare the gender ratios of displayed genders to that of participant genders. It is observed that men are shown a lot more to women than to men, and vice versa in the case of women as the displayed gender. We can attribute more men being shown to the skew in the number of male users to the female user on the application. While this distribution cannot be attributed to the algorithm alone, it raises the question of whether it would affect the experiences of people on the dating application, especially people who are seeking both Men and Women.

Of the people we interviewed, the bisexual women and the non-Binary person claimed that there was a huge disparity in the number of men and women shown to them. Even when filtering for only women, these users got male profiles in their feeds. The interviewees claim that these men set their gender as female and search for women in hopes of getting matches. This ruins the experience of other users and even causes them to take a break from using the application. One of the interviewees says about this unusual behavior, "When you turn on everyone it's mostly men being shown - it probably is mostly men because like, there's a lot of men on bumble, yes, but also a lot of men were marked as women are probably shown; but because you have everyone turn on you don't notice it." When the filters are changed to include everyone, the number of men only increases. On the other hand, the bisexual men we interviewed did not seem to notice such a huge disparity. While they were shown more men than women on average, they did not feel the difference was too big. This falls in line with our analysis results, wherein men are shown more to women and vice versa.

Aside from this disparity in profiles shown, there appears to be a disparity between the filters set by the user and the actual profiles that appear on their feed. One of our interview subjects reported seeing many middle-aged men in his feed despite setting the maximum age filter as 23. Bumble gives users the option to prevent profiles from slightly outside their filter range from showing up in their feed. However, even with this option enabled, most of our interview subjects felt that the male profiles that showed up on their feed were outside what they were looking for. One of the interviewees claimed, "I feel creepy when, I see the 'potential match missed' on a 30–40-year-old man."

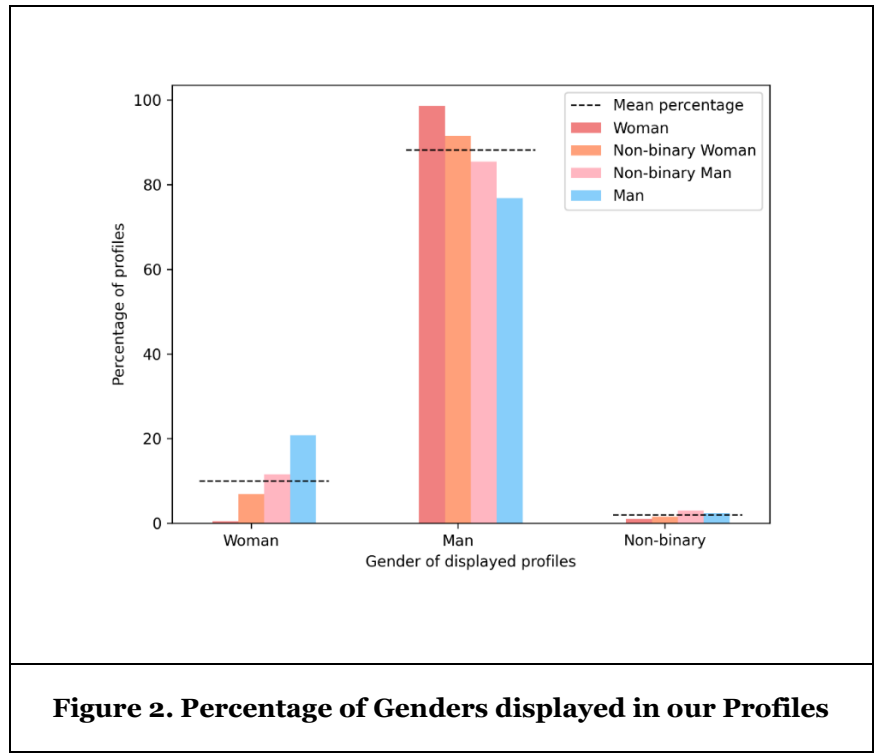

**Figure 2. Percentage of Genders displayed in our Profiles**

## Conclusion

Dating apps like Bumble utilize algorithms that learn from user data to make recommendations. As inferred from our findings, the algorithm falls prey to the existing biases in the datasets it has learnt on. If the datasets contain biases that are socially ingrained by a history of human and cultural prejudice [Raghavan et al., 2019], the algorithm can inadvertently perpetuate and amplify them, affecting the representation and experiences of users with diverse gender identities. Our study here, even though still in progress, has found potential biases and this stands in contrast to the present-day need for more inclusive platforms. It also provides a base for more studies which deal with many of the previously mentioned confounding variables, or to work towards a solution for more inclusive designs in dating applications. The technology behind dating apps like Bumble ensues the co-situation of individual social identities in a digitally enabled space. Even though the algorithm itself does not have intentions or consciousness, it is part of a system created and maintained by humans, who have moral responsibilities. Most of the current users on Bumble in India are its future citizens trying to look for romantic relationships. If Bumble's algorithmic bias filters out certain sections of the minority population and marginalizes some of them, it is reproducing dominant perceptions and constraining India's youth to enter liberal, progressive and open relationships. The recommendations algorithm, while designed to complement user preferences in dating, in turn, influences not just their dating behavior but also their social perception of other individuals. We attribute the moral agency of such systems to be compositely derived from algorithmic mediations and the design and utilization of these platforms. Our study not only highlights the discrepancies and potential biases in the algorithm but also serves as a catalyst to foster more inclusive design principles, thereby acknowledging the diverse needs of the current generation. From a business perspective, employing inclusive design practices can offer a competitive edge, in an overcrowded market. However, the risk of perpetuating biases is significant if algorithms are not programmed conscientiously, potentially leading to discrimination based on race, gender, or other traits [Vella, 2021]. From a technological standpoint, refining algorithms to be more inclusive of diverse gender identities and preferences can be implemented through machine learning techniques that are already integral to these platforms. This could involve training algorithms on more diverse datasets to reduce biases, an approach that aligns with the ethical and social responsibility aspects increasingly valued in technology.

By actively working against the ingrained societal biases in the algorithms, businesses can appeal to a broader and more diverse user base, while fostering an environment which reflects the progressive thoughts of the youth. This approach can potentially transform not just dating behaviors but also social perceptions, contributing to more liberal and open relationships among India's future citizens.